\documentstyle[prd,aps,psfig,floats]{revtex}
\def\b{\begin{equation}}
\def\e{\end{equation}}
\def\br{\begin{eqnarray}}
\def\er{\end{eqnarray}}
\def\l{\left}
\def\r{\right}
\def\pa{\partial}

\begin{document}
\draft
\title{Vanishing of cosmological constant in nonfactorizable geometry }
\author{T. Padmanabhan and S. Shankaranarayanan}
\address{IUCAA, Post Bag 4, Ganeshkhind, Pune 411 007, INDIA.\\
email:paddy@iucaa.ernet.in, shanki@iucaa.ernet.in}
\maketitle
\begin{abstract}
We generalize the results of Randall and Sundrum to a wider class of
four-dimensional space-times including the four-dimensional Schwarzschild 
background and de Sitter universe. We solve the equation for graviton 
propagation in a general four dimensional background and find an explicit 
solution for a zero mass bound state of the graviton. We find that this zero
mass bound state is normalizable only if the cosmological constant is 
{\it strictly} zero, thereby providing a dynamical reason for the vanishing
of cosmological constant within the context of this model. We also show that 
the results of Randall and Sundrum can be generalized without any modification
to the Schwarzschild background.

\end{abstract}
\vspace{0.05cm}
\section{Introduction}
Any realistic theory of gravity should be able to reproduce $r^{-1}$
behavior of the gravitational potential in the Newtonian limit. Generically, 
the potential falls off like $r^{-d + 3}$, where $d$ is the number of 
extra dimensions with infinite extend. Thus, to obtain $r^{-1}$ behavior, 
higher dimensional theories of gravity have been assuming compactness 
($\approx$ Planck length) of the extra dimensions. Thus, in the conventional 
way of extracting an effective lower-dimensional theory from higher 
dimensions, one performs a Kaluza-Klein reduction in which the extra 
dimensions are warped up into a compact space (of the order of Planck length) 
such as a torus or a sphere[see Ref.~\cite{review} and references there in]. 
Provided that the scale size of these internal dimensions is 
sufficiently small in relation to the energy scale of excitations in the 
lower dimension, then the mass gap separating the massless modes from the 
massive modes will be sufficient 
to ensure that the internal dimensions are essentially unobservable, 
and the world will essentially appear to be effectively lower dimensional. 
If an extra dimension is non-compact, there would be a continuum modes with 
masses extending down to zero, when seen from the lower-dimensional 
view-point. 
\par
Recent developments in string theory have shown that if
matter fields are localized on a 3-brane in $1 + 3 + d$ dimensions, while 
gravity can propagate in the extra dimensions, then the extra dimensions can be
large \cite{arkani,shiu}. In this scenario, the Planck scale $M_P$ is traded 
for the size of the extra dimensions felt by gravity. Likewise, gauge coupling
unification can be preserved and remain perturbative, but now occurs at scales
as low as a TeV. One can therefore have gravity and gauge coupling unification
occurring at as low a scale as a few hundred GeV to 1 TeV. This new scenario 
has  been claimed to be experimentally testable \cite{che} and  offers a 
simple qualitative explanation to the fermion mass hierarchy problem 
\cite{ransu1}.
\par
In these large extra spatial dimensions, deviations from Newtonian potential 
will be detected at the scale of the extra dimensions. The 
form of the Newtonian potential can be obtained for a point-like mass, 
in these models, by means of Gauss' law \cite{arkani}. Denoting by $r$ the 
radial distance in $4+d$ dimensions and by $r_b$ the radial distance as 
measured on the 3-brane, we find for distances $r$ much greater than the 
typical size of the extra dimension $L$ a potential of the form 
\b
V_{(4)}=-G_N\,{M\over r_b}
\ ,
\label{V>}
\e
where $G_N=m_p^{-2}$ is Newton's constant in four dimensions.
On the other hand for $r\ll L$ the potential becomes
\b
V_{(4+d)}=-G_{(4+d)}\,{M\over r^{1+d}}
\ ,
\label{V<}
\e
with $G_{(4+d)}=M_{(4+d)}^{-2-d}=L^d\,G_N$.
This implies that the huge Planck mass
$m_p^2=M_{(4+d)}^{2+d}\,L^d$ and, for sufficiently large
$L$ and $d$, the bulk mass scale $M_{(4+d)}$ (eventually
identified with the fundamental string scale) can be as small
as $1\,$TeV.
Since
\b
L\sim \left[{1\,{\rm TeV}/M_{(4+d)}}\right]^{1+{2\over d}}\,
10^{{31\over d}-16}\,{\rm mm}
\ ,
\label{tev}
\e
demanding that Newton's law is not violated for distances larger
than $1\,$mm restricts $d\ge 2$ \cite{arkani,long}.
Further bounds are obtained by estimating the production of
Kaluza Klein gravitons and support higher values of $d$ \cite{bounds}.
\par
On the other hand, Randall and Sundrum(RS, hereafter) \cite{ransu2} have 
shown that these extra dimensions in five-dimensional space-times need 
not be compact. They have shown that for $d = 1$, gravity can be localized 
on a single 3-brane (where the standard model particles are confined) even 
when the fifth dimension is infinite. The non-compact localization arises 
via the exponential warp/conformal factor in the nonfactorizable metric: 
\b
ds^2 = \exp(-2 k |y|) \l[ dt^2 - d{\bf x}^2\r] - dy^2. 
\label{eq:rs1}
\e
The metric signature we adopt is $(+ - - - -)$. For $y \neq 0$, this 
metric satisfies the five dimensional Einstein's equation with 
negative five dimensional cosmological constant, $\Lambda \approx - k^{2}$. 
The brane is located at $y = 0$, and the induced metric on the brane is a
Minkowski metric. The bulk is a five dimensional anti-de Sitter metric, with 
$y =0$ as boundary, so that $y < 0$ is identified with $y > 0$, reflecting the
$Z_2$ symmetry with the brane as fixed point, that arises in the string theory.
\par
Perturbations of the metric (\ref{eq:rs1}) shows that the Newtonian potential 
on the brane is recovered at lowest order:
\b
V(r) = \frac{G M}{r}\l(1 + \frac{2 }{3 k^2 r^2}\r).
\e
Thus, the four dimensional gravity is recovered at high energies, with a
first-order correction that is constrained by current submillimeter 
experiments~\cite{hoyle}. The zero mode produces the standard $1/r$ 
gravitational potential along the brane, and the Kaluza-Klein modes give 
rise to corrections of order $1/r^3$. [The general line element of the form 
in Eq.~(\ref{eq:rs1}) has been obtained earlier by 
Gogberashvili~\cite{gog} by setting the momentum toward the large extra, 
fifth, dimension to be zero.] 
\par
The corrections to the Newtonian gravitational potential $V_N(r)\propto 
(m_1 m_2/r)$ have been investigated earlier by several authors from 
different points of view. Duff \cite{duffold} had obtained a similar 
result by computing the one-loop corrections to the (flat) graviton propagator. 
In his analysis, the single graviton exchange provided the linearized 
Schwarzschild line element, which in the weak field limit is the standard 
$1/r$ potential and the inclusion of  the quantum corrections to one-loop 
order modifies gives rise to corrections of order $1/r^3$. Since the 
lowest order corrections have to be linear in $G\hbar$, it is obvious 
from dimensional grounds that the correction will multiply $V_N$ by a 
factor of the form $[1+a(G\hbar/c^3r^2)]$ where $a$ is numerical 
coefficient. [While this is the leading {\it quantum} correction, it 
may be noted that the lowest order post-Newtonian approximation will give a 
correction of the form $[1+b(G(m_1+m_2)/c^2r)]$, where $b$ is a numerical 
factor, which has a slower fall-off with distance].
\par
Danoghue~\cite{donomg} has obtained similar results by treating gravity as an 
effective field theory. He argues that the leading quantum corrections, in
powers of the energy or inverse powers of the distance, can be computed in
quantum gravity through the knowledge of the low-energy structure of the
theory (effective field theory). He shows that the one loop corrections to 
the graviton propagator gives the $1/r^3$ corrections to the Newtonian
potential. He also emphasizes that the correction to low energy 
gravity, treated as an effective theory is remarkably unique and the 
leading quantum correction to the potential is $(1/r^3)$.
[There have been other similar analysis in the literature where 
the classical and quantum corrections to the Newtonian potential have been
calculated. See for example, Ref.~\cite{longrange}]
\par
In the case of RS, there is no background Schwarzschild metric 
and they merely study the graviton perturbations around the {\it flat} 
four-dimensional spacetime. Their approach is essentially to look at the 
corrections to the graviton propagator arising from a set of continuum states 
with mass $m>0$. The analysis by itself is classical and indeed, the 
corrections to $V_N$ which they find does not depend on $\hbar$ directly; of 
course, they provide an {\it interpretation} which is quantum mechanical. In 
contrast, much of the earlier work, concentrated on quantum gravitational 
corrections, have used the background Schwarzschild line element.
\par
This raises the question: Is it possible to generalize the ideas of RS to
a situation in which the four-dimensional metric is nontrivial (say, a 
Schwarzschild metric or de Sitter universe)? Will we get the same mass 
spectrum for the graviton modes and the same correction term to $V_N(r)$ ? The 
fact that Duff and Danoghue obtained similar results suggests that this could 
be the case -- though it needs to be explicitly demonstrated. 
\par
In this paper, we show that the main results of RS have a 
simple mathematical origin and can indeed be generalized to a wider class of 
models. We will  provide a {\it general} solution to the zero mass graviton 
mode in {\it arbitrary} background and --- as an illustration ---  will work 
out explicitly the case that incorporates a spherically symmetric solution in 
four dimensions. (This will include as special cases, the Schwarzschild and de 
Sitter manifolds). It is important to show that the properties of the graviton 
propagation, and the effective gravitational potential does not change under 
such a generalization. We shall provide exact solutions which demonstrate that 
such is indeed the case; these solutions also provide some insight into the 
structure of the solution and will possibly allow us to study --- for example 
--- models for black hole evaporation in this context.
\par
In Section (II), we will solve the equation for graviton propagating in 
general four dimensional space-time and obtain an explicit solution for the 
zero mass bound state of the graviton. In section (III), we perform the 
analysis for the four dimensional spherically symmetric space-times and show 
explicitly that the four dimensional cosmological constant should vanish. 
Finally in section (IV) we summarize the results and discuss the implication 
of the result in the compactified Randall-Sundrum model.


\section{Generalization of Randall-Sundrum model}

In this section, we study the plane wave gravitons, $h_{\mu\nu}$, 
propagating in the five dimensional space-time, 
\b
ds^2 = g_{ab} dx^a dx^b
= \exp[-2 a(y)]\l[g^{(4)}_{\mu\nu}dx^\mu dx^\nu\r] - dy^2,
\label{eq:rsext1}
\e
with the condition that it satisfies the full five dimensional Einstein's
equation with the five dimensional cosmological constant. We use the lowercase 
Latin letters for the full five dimensions and the lowercase Greek letters 
for four dimensions. [We follow the notation of RS closely to provide 
easy comparison.]
\par
Denoting the perturbed metric by $\tilde{g}_{ab} = g_{ab} + h_{ab}$ and using 
the gauge 
\b
h_{55} = h_{5\mu} =0, ~~~\nabla^{\mu} h_{\mu\nu} = 0, ~~~h^{\mu}_{\mu} =0,
\e
it is easy to see that $h_{\mu\nu}$ can be written as plane wave 
gravitons, {\it i.e.} $h_{\mu\nu} = e_{\mu\nu} \Phi$  where $e_{\mu\nu}$ is 
the polarization tensor. The equation satisfied by $\Phi$ can
be separated with the ansatz $\Phi(x^\mu,y)=A(x^\mu)Z(y)$. Substituting into
the wave equation, separating the variables using a constant $m^2$, we find
that $A$ satisfies the standard wave equation for a particle of mass $m$ while
$Z$ satisfies the equation:
\b
\frac{d^2 Z}{dy^2} + \l( - 4 \dot{a}^2(y) + 2 \ddot{a}(y)
+ m^2 \exp[2 a(y)] \r)Z = 0,
\label{schr1}
\e
[The essential steps leading to the above equation is been given in Appendix 
A.] This reduces to equation (8) of RS, when we use their solution 
$a(y)= k|y|$. We are interested in the allowed range of values for $m$ and 
whether we can get an acceptable solution for $m=0$. By inspection, it is 
clear that this equation has a solution for $m=0$, given by
\b
Z=\exp[-2 a(y)].
\label{schr2}
\e
In fact, this is {\it precisely} the ground state wave function which RS obtain 
(after a series of algebraic transformations! ) for their special case of 
$a(y)=k|y|$. The physical meaning, mathematical simplicity and generality 
of the result is hidden by: (i) their transformations and (ii) the fact that 
they never give $\psi(y)$ but only $\hat\psi(z)$ in their paper. [Note that 
Eq.~(\ref{schr2}) is a valid solution to Eq.~(\ref{schr1}) with $ m = 0$ a long
as $a(y)$ is continuous even if its derivative is discontinuous at the origin.]
\par
This is the first result of this paper and shows that the existence of a 
zero mass graviton is a very general result and does not require much of 
the extra assumptions in RS except that $Z$ should be well 
behaved and {\it normalizable} as a function of $y$, in the relevant range. 
[Note that the ground state wave function for an arbitrary four dimensional 
line element is exactly the conformal/warp factor in the generalized 
Randall-Sundrum model.] This clearly shows that the stability of the 
3-brane can be explicitly shown in the RS model by obtaining the zero mass 
graviton wavefunction which is well behaved and normalizable. The question 
arises 
as to the conditions under which we will obtain a normalizable function 
for $Z(y)$. Such an analysis for a general $a(y)$ is complicated and hence 
we will illustrate it explicitly for a special case. In the next section, 
we take a simple case by assuming that the four dimensional spacetime is 
spherically symmetric and show that for the case of non-zero four dimensional 
cosmological constant, the zero mass ground state wave function is 
non-normalizable. 
\par
We would also like to point out the following: The other eigen-values and 
eigen functions can be found by converting Eq.~(\ref{schr1}) into an 
eigenvalue equation for $m^2$. In general, an equation of the form
\b
\frac{d^2S}{dx^2} + \l( E~ V(x) - 4k^2 \r)S = 0,
\label{app1}
\e
(where $E$ and $k^2$ are constants, $V(x)$ is a continuous function of $x$)
can be transformed to an eigenvalue equation for $E$ by changing the
independent variable from $x$ to $z$ by  
\b
z = \int dx V(x)^{1/2}
\label{trans}
\e
and dependent variable from $S$ to $\hat{S}=S V^{-1/4}$. This will give a
modified Schroedinger equation of the form
\b
\frac{d^2\hat{S}}{dz^2} + \Bigg[- \frac{1}{16} \l(\frac{d(\ln(V(x))}{dz}
\r)^2 + \frac{4k^2}{V(x)} - \frac{1}{4} \frac{d^2(\ln(V(x))}{dz^2} \Bigg] 
\hat{S} = -E \hat{S}
\label{app12},
\e
where, $x$ in the above expression is expressed in terms of $z$ using
Eq.~(\ref{trans}).
\par


\section{Special Case: Spherically Symmetric space-time} 

In the previous section, we have shown that the existence of the zero mass 
graviton is a very general result in the case of RS model. However, the 
analysis of the normalization of (zero mass) ground state wavefunction for 
a general four dimensional space-time is complicated. Here, we take a 
simple case by assuming that the four dimensional spacetime is spherically 
symmetric and is of the form 
\b
ds^2 = \exp(-2 a(y))\l[A(r) dt^2 - B(r) dr^2 - r^2 d\Omega^2 \r] - dy^2,
\label{rsext}
\e
\noindent where, $d\Omega^2$ is the angular line element and $a(y)$, $A(r)$ 
and $B(r)$ need to be determined {\it via} the five-dimensional Einstein's 
equations. We consider the latter to be of the form 
\b
G_{ab} = \Lambda g_{ab}
\e
with possible non-zero vacuum energy density $\Lambda$ in five dimensions. 
Inserting the ansatz (\ref{rsext}) for the metric, the only non-vanishing 
components of the Einstein tensor, $G$, are the diagonal components. The 
Einstein's equation, for $(00)$ and $(11)$ components, reduces to 
\br
& &{\frac{1}{r^2}} - {\frac{1}{r^2 B(r)}} + {\frac{B'(r)}{r B^2(r)}}
= \exp[-2 a(y)] R(y)
\label{modg00} \\
&-&{\frac{1}{B(r)}}\l [{\frac{1}{r^2}} - {\frac{B(r)}{r^2}} + {\frac{A'(r)}
{r A(r)}}\r ]
= \exp[-2 a(y)] R(y)
\label{modg11}
\er
where,
\b
R(y) = \Lambda + 6 \dot{a}^2(y) - 3 \ddot{a}(y),
\e
\noindent and the prime denotes derivative with respect to $r$. Combining the 
two equations, we obtain $B(r) = 1/A(r)$. Substituting for $B(r)$ in the above
equations and to the $(22)$ and $(33)$ components of the Einstein's equation,
we get
\br 
-A(r) \l( {\frac{1}{r^2}} - {\frac{A'(r)}{r A(r)}} - {\frac{1}{r^2 A(r)}} \r )
&=&\exp[-2 a(y)] R(y) 
\label{fing00} \\
-{\frac{1}{2 r}} \l( 2 A'(r) + r A''(r) \r) &=&
\exp[-2 a(y)] R(y)  
\label{fing22} \\
-\l({\frac{A''(r)}{2}} + 2 {\frac{A'(r)}{r}} + {\frac{A(r) - 1}{r^2}}\r)
&=&\exp[-2 a(y)] \l ( \Lambda + 6 \dot{a}^2(y) \r ). 
\label{fing55}
\er
Solving the above equations, gives $A(r)$ to be 
\b
A(r) = 1 -\frac{C}{r} - \frac{\lambda}{3} r^2
\label{Ar},
\e
where $C$ and $\lambda$ are the constants of integration. This four-dimensional 
metric is the well known Schwarzschild - de Sitter metric for the choice of 
$C > 0$, where $\lambda$ is the four-dimensional cosmological constant, in the
sense that the four dimensional metric with $A(r)$ given by Eq.~(\ref{Ar})
corresponds to a four dimensional space-time with this cosmological constant.
[We use the term cosmological constant in four dimensions in the above sense and
it should not be confused with the other possible ways of defining the
cosmological constant --- for example, from the brane tension {\it etc.} Note 
that the sign of $\lambda$ is still undetermined.] Substituting the form of 
$A(r)$ in the original equations, the differential equation for $a(y)$ becomes 
\b
\frac{d^2 a(y)}{d y^2} = \frac{\lambda}{3} \exp(2 a(y)).
\label{ay}
\e
It is clear that the conformal factor will have {\it only} the $\lambda$
dependence and will be independent of $C$. [Normally the four dimensional
space-time can have a non-vanishing cosmological constant only when there is a
source in the right hand side of the four dimensional Einstein's equations. In
our case, if we write the five dimensional $G_{ab}$ in terms of four
dimensional Einstein tensor $G_{\mu\nu}$ and extra terms arising from the 
fifth dimension, it is possible to show that the effective source for 
$G_{\mu\nu}$ is exactly that corresponding to a four dimensional cosmological
constant $\lambda$.]
\par
Solving Eq.~(\ref{ay}), it is easy to obtain the form of $a(y)$ such that 
it reduces to the RS result of $a(y) = k |y|$ when $\lambda =0$. We get
\b
\!\!\exp[- 2 a(y)] =  \exp[2 k |y|] \l[\exp[- 2 k |y|]-
(\lambda/12k^2) \r]^2 \!\!
\label{apsi}
\e
with $k$ being a constant related to $\Lambda$ by $\Lambda=-6k^2$. This shows 
that $\Lambda<0$ for an acceptable solution. Equations (\ref{apsi}), 
(\ref{Ar}) with the result $A(r) = 1/B(r)$ completely determine the metric. The
modulus sign in $|y|$ will make the derivatives of $a(y)$ discontinuous at the
origin $ y = 0$ which can be taken to be the location of the membrane as in the
RS case.
\par
Eq.~(\ref{apsi}) allows us to draw an important conclusion which is the second 
key result of this paper. Note that the conformal factor 
$Z = \exp[-2 a(y)]$ depends on $\lambda$ but {\it not} on $C$. In the limit 
of $\lambda \rightarrow 0$ the conformal factor for the four-dimensional 
world line element is same as in the RS model. Thus, the original analysis of 
RS can be generalized {\it without any modifications} to the case in which the 
four dimensional spacetime is described by Schwarzschild line-element 
[$\lambda =0$, $C >0$ in equation (\ref{Ar})] as well suggesting that the 
zeroth order gravitational interaction, in the form of Schwarzschild line 
element, gets ``corrected" by the conformal factor. This could possibly be the 
reason why the one-loop corrections to Schwarzschild metric in the earlier 
analysis of Duff~\cite{duffold} also gives similar result. 
\par
The above reason is strengthened by the results in Ref.~\cite{duff,alva}: 
In a recent paper, Duff \cite{duff} has shown that the propagator for the 
continuum graviton modes, in the RS picture, incorporates all quantum effects 
of matter on the brane. Using the Duff's analysis, Alvarez and Mazzitelli
\cite{alva} have shown that the for conformal fields and up to quadratic 
order in the curvature, the non-local effective action is equivalent to the 
$d + 1$ action for classical gravity in $AdS_{d + 1}$ restricted to a 
$d - 1$ brane. 
\par
The condition on the {\it four-dimensional } cosmological constant $\lambda$ is 
more interesting. The ground state wave function $Z = \exp(-2 a(y))$ in 
(\ref{apsi}) is not normalizable for $\lambda \neq 0$ and hence we do not get a
massless [$m = 0$ in equation (\ref{schr1})] graviton for $\lambda
\neq 0$. An examination of the general solution to (\ref{ay}) confirms this 
conclusion. Using $Z = \exp(-2 a(y))$, the first integral to (\ref{ay}) can be 
written as:
\b
\frac{dZ}{dy} = \pm \l(4 \beta_1 Z^2 + \frac{4 \lambda}{3} Z\r)^{1/2},
\e
where $\beta_1$ is the constant of integration. For $\beta_1 < 0$, the 
solution is oscillatory with nodes and hence is not of interest. For the case 
$\beta_1 = k^2 > 0$, we obtain the solution to be 
\b
Z = - \frac{\lambda}{6 k^2} + \frac{1}{16 k^2} \exp(\pm 2 k (y - y_0)) 
+ \frac{\lambda^2}{9 k^2} \exp(\mp 2 k(y - y_0)),
\label{fin}
\e
where $y_0$ is the constant of integration. In the case of $\lambda = 0$, the 
wave function($Z$) is normalizable and reduces to the ground state wave
function obtained by RS with a suitable choice of the signs for $y >0$ and 
$y < 0$ [We take the solution to be varying as $\exp(-2 k y)$ for $y > 0$ and 
$\exp(2 k y)$ for $ y < 0$ with the membrane being located at $ y = 0$].
However, when $\lambda \neq 0$, the wave function is not bounded as 
$|y| \to \infty$ (for any combination of signs in the argument of the
exponential) and hence is not normalizable for non-zero $\lambda$. This is
because the third term on the right hand side of Eq.~(\ref{fin}) (which is
non-zero when $\lambda \neq 0$) comes with an argument to the exponential
having a different sign compared to the second term. This shows clearly that
the nature of the solution for $Z(y)$ --- which acts as the ground state
wave-function for zero mass graviton mode --- is very different when $\lambda
\neq 0$ compared to the case considered by RS. [The above result can be 
understood in a slightly different manner: 
The ground state wave function in Eq.~(\ref{schr2}) is the same as the 
conformal factor of the line-element (\ref{eq:rsext1}). If the ground 
state wave-function blows up as $y \to \infty$ then the conformal/warp 
factor in the Randall-Sundrum line element will be very large for large 
$y$. Hence, the brane located at $y = 0$ is unstable to the metric
perturbations.]
\section{Conclusions and Discussions}

To conclude, we have shown that the existence of a zero graviton mode is 
general, i.e. it exists for a wide class of four dimensional metrics in the
case of RS model. In particular, the results of RS are valid without 
modifications for a four dimensional Schwarzschild black hole. But the 
presence of non-zero cosmological constant in four dimensions modifies the 
RS results. The presence of a non-zero cosmological constant does not 
provide a normalizable ground state wave
function corresponding to the zero mass graviton. Hence, we have obtained a 
dynamical reason for the strict vanishing of the cosmological constant 
{\it within the context of these models}. The stability of the 3-brane to 
different classes of matter fields in the context of the general five
dimensional metric is under investigation. 
\par
We would like to point the reader the difference in the approach taken here 
and to the earlier works~\cite{sch_brane}: 
%
%
The earlier analysis of the Schwarzschild metric on the brane were 
performed by taking the case $a(y) = k |y|$. In this case, it is easy to
demonstrate that the ${\hat R}_{ab}$ solves the RS equations of 
motion, provided the four dimensional brane is Ricci flat ($R_{\mu\nu} = 0$). 
Hence in these analysis, replacing the Ricci flat branes with the flat 
branes was by forcing the conformal/warp factor to be same as that of RS.
\par
Our analysis in this paper is geared towards understanding the 
stability of the 3-brane against the metric perturbations (in the five 
dimensions) for a general four dimensional space-times. We have shown that 
the stability of the 3-brane in the RS model can be explicitly shown in the 
RS by obtaining the zero mass ground state graviton wavefunction which is 
well behaved and normalizable. Here we have performed this analysis for a 
four dimensional spherically symmetric metric and obtained the general form 
of the four dimensional spherically symmetric metric 
along with the conformal factor $a(y)$ by solving the five dimensional 
vacuum Einstein's equation (with non-zero $\Lambda$). The general solution 
we obtained shows that there the conformal/warp factor is independent of the 
Schwarzschild mass [see section (III)]. However, the analysis (of the four
dimensional Schwarzschild metric in the 3-brane) by earlier authors is by 
forcing the conformal/warp factor to be same as that of RS and hence replacing
the Ricci flat branes with the flat branes. The reason for conformal factor to 
be independent of the Schwarzschild mass [constant $C$ in Eq.~(\ref{Ar})] is 
not clear in the earlier works.  
\par
An interesting alternative scenario would be to use the model by
RS in Ref.~\cite{ransu1}. In this scenario, we can set up two 3-branes where
the 3-branes are extended in the $x_{\mu}$ directions and are located at some
fixed points in the $y$ axis and thus restricting the extra dimensions to be
compactified. [In this model, it is assumed that the branes do not contribute 
to the energy momentum tensor.] By restricting the extra dimensions to be 
compactified, we can obtain normalizable zero mass gravitons. Such an analysis 
leads to two different situations depending on whether: (i) $\lambda > 12k^2$ 
or (ii) $\lambda < 12 k^2$. The first possibility, even if $k \approx TeV$, 
will give a large cosmological constant. The other case, which is more 
plausible, gives us the upper bound on the compactification scale(radius) 
of the extra dimensions. [Some of these issues have been discussed in
Ref.~(\cite{inf}).] These issues are under current investigation. 
\par
Finally, we would like to mention the following curious fact: In
conventional four-dimensional general relativity, linearizing 
the Einstein-Hilbert action,
\b
S_{gravity} = -\frac{c^3}{16\pi G}\int \sqrt{-g(x)}\,
\l[ R(x) + 2 \lambda \r]\, d^4 x
\e
(where $R(x)$ is the Ricci scalar, $\lambda$ is the cosmological
constant and $g_{\mu\nu}$ is the general four-dimensional metric),
we obtain,
\b
\Box^{(4)} h_{\mu\nu} = -\lambda h_{\mu\nu}.
\label{eq:lin}
\e
The cosmological constant appears as a mass term in the linearized spin-2 
wave equation. Vanishing of cosmological constant is required for 
this equation to be interpreted as representing the massless spin-2 
particles (gravitons) in general. [The graviton propagation in de Sitter 
background (which is a maximally symmetric space-time) has been performed 
(see for example Ref. \cite{allen}) and it was shown that gravitons 
possesses only two physical propagating degrees of freedom. A detailed
analysis for a {\it general} background has not been performed and in these 
cases the vanishing of the cosmological constant is required to interpret 
it as representing massless gravitons (corresponding to a long range 
interaction).] However, by making the
cosmological constant to be very small one can obtain a long range
interaction for gravity. Our analysis here shows that even {\it an 
arbitrarily small} cosmological constant will make the ground state
wave-function (corresponding to a massless graviton) to be non-normalizable, 
requiring the cosmological constant to strictly vanish. 
Whether there exists a deeper connection between the two results is not 
clear and is under investigation.  

\section*{\centerline{ACKNOWLEDGMENTS}}
\noindent
We thank Naresh Dadhich for fruitful discussions and for drawing
us into the fifth dimension. We thank K.~Subramanian for comments on the
earlier draft of the paper. S.S. is being supported by the Council of 
Scientific and Industrial Research, India.
\appendix
\section{}
For the sake of completeness, we outline the essential steps leading to 
Eq.~(\ref{schr1}) in section (II). Defining 
$\Theta^a_{\mu\nu} \equiv g^{ac}h_{\mu\nu;a}$ (The semicolon on the right hand
side represents the covariant derivative), we 
have
\br
\nabla^a \nabla_a h_{\mu\nu} = g^{ac} \nabla_c\l(h_{\mu\nu;a}\r) 
&=& \Theta^c_{\mu\nu ;c} \nonumber \\
\label{eq:def_the1}
&=& \Theta^c_{\mu \nu ,c} + \Theta^m_{\mu\nu}\Gamma^c_{mc} - \Theta^c_{m\nu} 
\Gamma^m_{c \mu} - \Theta^c_{m\mu} \Gamma^m_{c\nu} \\
\label{eq:def_the2}
&=& (- g)^{-1/2} \pa_c\l(\sqrt{-g} \Theta^c_{\mu\nu}\r) 
- \Theta^c_{m\nu} \Gamma^m_{c \mu} - \Theta^c_{m\mu} \Gamma^m_{c\nu} 
\er
Evaluating $\Theta$'s in the R.H.S of the expression (\ref{eq:def_the1}),
we obtain
\br
\label{eq:the1}
\Theta^c_{\mu\nu}&=& g^{ca} h_{\mu\nu ;a} \nonumber \\
&=& g^{ca} \l[ \pa_a h_{\mu\nu} - \Gamma^{\sigma}_{\mu a} h_{\sigma \nu}
- \Gamma^{\sigma}_{\nu a} h_{\mu \sigma} \r] \\
\label{eq:the3}
\Theta^c_{m\nu}&=& g^{ca} h_{m \nu;a}\,=\, g^{ca} h_{\eta\nu;a} \qquad
({\rm using~the~gauge~condition}~h_{5\mu} = 0) \nonumber \\
&=& g^{ca} \l[ \pa_a h_{\eta \nu} - \Gamma^{\sigma}_{\eta a} h_{\sigma\nu}
- \Gamma^{\sigma}_{\nu a} h_{\eta \sigma}\r] \\
\label{eq:the4}
\Theta^c_{\mu m}&=& g^{ca} h_{\mu m;a}\,=\, g^{ca} h_{\mu \eta;a} \qquad
({\rm using~the~gauge~condition}~h_{5\mu} = 0) \nonumber \\
&=& g^{ca} \l[ \pa_ah_{\mu\eta} - \Gamma^{\sigma}_{\mu a} -
\Gamma^{\sigma}_{\eta a} h_{\mu \sigma}\r]
\er
We know
\br
\Gamma^i_{kl}&=& \frac{1}{2} g^{im} \l[ \frac{\pa g_{mk}}{\pa x^l}
+ \frac{\pa g_{ml}}{\pa x^k} - \frac{\pa g_{kl}}{\pa x^m} \r] \\
\Gamma^{\sigma}_{\mu a}&=&\frac{1}{2} g^{\sigma m} \l[ \frac{\pa g_{m\mu}}
{\pa x^a} + \frac{\pa g_{ma}}{\pa x^{\mu}} - \frac{\pa g_{\mu a}}{\pa x^m}\r]
\er
$\Gamma$ can be easily evaluated for the line element (\ref{eq:rsext1}) and is
given by
\b
\Gamma^{\sigma}_{\mu a} = \frac{1}{2} g^{\sigma \beta} \l[
\frac{i\pa g_{\beta\mu}} {\pa x^a} + \frac{\pa g_{\beta a}}{\pa x^{\mu}}
- \frac{\pa g_{\mu a}}{\pa x^{\beta}} \r] \delta^a_{\alpha} - 
\dot{a}(\psi) \delta^{\sigma}_{\mu} \delta^a_5.
\e
Thus, the Eqs.~(\ref{eq:the1},~\ref{eq:the3},~\ref{eq:the4}) will get modified 
to the form,
\br
\Theta^c_{\mu\nu}&=&g^{ca} \l[ \pa_a h_{\mu\nu} + 2 \dot{a}(\psi) \delta^a_5
h_{\mu\nu}\r] \nonumber \\
&-&\!\! \frac{g^{ca}}{2} \l[g^{\sigma\beta}\l(\pa_a\,g_{\beta\mu} + \pa_{\mu}\,
g_{\beta a} - \pa_{\beta}\, g_{\mu a}\r) \delta^a_{\alpha} h_{\sigma\nu} + 
g^{\sigma\beta}\l(\pa_a\,g_{\beta\nu} + \pa_{\nu}\,g_{\beta a} - 
\pa_{\beta}\, g_{\nu a}\r) \delta^a_{\alpha} h_{\mu\sigma}\r] \\ 
\Theta^c_{m\nu}&=& g^{ca} h_{\eta\nu;a}\,=\,
g^{ca} \l[ \pa_a h_{\eta\nu} + 2 \dot{a}(\psi) \delta^a_5 h_{\eta\nu}
\r] \nonumber \\ 
&-&\!\! \frac{g^{ca}}{2} \l[g^{\sigma\beta}\l(\pa_a\,g_{\beta\eta} +
\pa_{\eta}\,g_{\beta a} - \pa_{\beta}\, g_{\eta a}\r) \delta^a_{\alpha}
h_{\sigma\nu} + g^{\sigma\beta}\l(\pa_a\,g_{\beta\nu} + 
\pa_{\nu}\,g_{\beta a} - \pa_{\beta}\, g_{\nu a}\r) \delta^a_{\alpha}
h_{\eta\sigma}\r] \\
\Theta^c_{\mu m}&=& g^{ca} h_{\mu\eta ;a}\,=\,
g^{ca} \l[ \pa_a h_{\eta\mu} + 2 \dot{a}(\psi) \delta^a_5 h_{\eta\mu}
\r] \nonumber \\ 
&-&\!\! \frac{g^{ca}}{2} \l[g^{\sigma\beta}\l(\pa_a\,g_{\beta\mu} +
\pa_{\mu}\,g_{\beta a} - \pa_{\beta}\, g_{\mu a}\r) \delta^a_{\alpha}
h_{\sigma\eta} + g^{\sigma\beta}\l(\pa_a\,g_{\beta\eta} + 
\pa_{\eta}\,g_{\beta a} - \pa_{\beta}\, g_{\eta a}\r) \delta^a_{\alpha}
h_{\mu\sigma}\r] 
\er
We obtain the full expression of $h_{\mu\nu;a}^{;a}$ by substituting these
expressions obtained for $\Theta$ in Eq.~(\ref{eq:def_the1}).  The first 
term in the R.H.S of Eq.~(\ref{eq:def_the1}) is
\br
(- g)^{-1/2} \pa_c\l(\sqrt{-g} \Theta^c_{\mu\nu}\r)&=&
\Box^{(4)} h_{\mu\nu} + \pa_5^2 h_{\mu\nu} - 2 \dot{a}(\psi) \pa_5 h_{\mu\nu}
- 8 \dot{a}^2(\psi) h_{\mu\nu} + 2 \ddot{a}(\psi) h_{\mu\nu} \nonumber \\
&-& \frac{(- g)^{-1/2}}{2} \pa_{\tau}\l[ \sqrt{-g} g^{\tau\alpha}
g^{\sigma\beta} \l(\pa_{\alpha}g_{\beta\mu} + \pa_{\mu}g_{\beta\alpha}
- \pa_{\beta}g_{\mu\alpha}\r) h_{\sigma\nu}\r] \nonumber \\
&-& \frac{(- g)^{-1/2}}{2} \pa_{\tau}\l[ \sqrt{-g} g^{\tau\alpha}
g^{\sigma\beta} \l(\pa_{\alpha}g_{\beta\nu} + \pa_{\nu}g_{\beta\alpha}
- \pa_{\beta}g_{\nu\alpha}\r) h_{\mu\sigma}\r],
\label{eq:term1}
\er
where $\Box^{(4)}$ denotes the four dimensional D'Alembertian operator. The
second and third terms in the R.H.S of the expression (\ref{eq:def_the1}) are
\br
\label{eq:term2}
\Theta^c_{\eta\nu} \Gamma^{\eta}_{c\mu}&=& \frac{1}{2}\, III_a \,
g^{\tau\alpha} \pa_{\alpha}h_{\mu\nu} - \dot{a}(\psi) \pa_5\,
h_{\mu\nu} - 2 \dot{a}^2(\psi) r\pa_5\, h_{\mu\nu} \nonumber \\
&-& \frac{g^{ca}}{4} \l[I_a \times III_a \times \delta^a_{\alpha} h_{\sigma\nu}
+ II_a \times III_a \delta^a_{\alpha} h_{\eta\sigma}\r] \\
\label{eq:term3}
\Theta^c_{\mu\eta} \Gamma^{\eta}_{c\nu}&=& \frac{1}{2}\, III_b\,
g^{\tau\alpha} \pa_{\alpha}h_{\mu\nu} - \dot{a}(\psi) \pa_5\,
h_{\mu\nu} - 2 \dot{a}^2(\psi) \pa_5\, h_{\mu\nu} \nonumber \\
&-& \frac{g^{ca}}{4} \l[I_b \times III_b \times \delta^a_{\alpha}\,
h_{\sigma\nu} + II_b \times III_b \times \delta^a_{\alpha}\, h_{\eta\sigma}\r]
\er
Where $I_{a,b}\, ,\, II_{a,b} \, , \, III_{a,b}\,$ are the terms which depend
only on the four-dimensional coordinates $X^{\mu}$. Combining the terms in the
expressions (\ref{eq:term1},~\ref{eq:term2},~\ref{eq:term3}) and rearranging 
them, we get
\br
h_{\mu\nu;a}^{;a}&=& \Box^{(4)} h_{\mu\nu} + \pa_5^2 h_{\mu\nu} -
4 \dot{a}^2 h_{\mu\nu} + 2 \ddot{a} h_{\mu\nu} \nonumber  \\
&+& {\rm terms~depending~on~the~4-D~coordinates}
\label{eq:fin_h}
\er
Thus, it is easy to see from the above relation that $h_{\mu\nu}$ can be
written as plane wave gravitons {\it i.e.} $h_{\mu\nu} = e_{\mu\nu} \Phi$.
The equation satisfied by $\Phi$ can be separated with the ansatz 
$\Phi(x^\mu,y)=A(x^\mu)Z(y)$. Substituting into 
the wave equation, separating the variables using a constant $m^2$, we find
that $A$ satisfies the standard wave equation for a particle of mass $m$ while
$Z$ satisfies the equation:
\b
\frac{d^2 Z}{dy^2} + \l( - 4 \dot{a}^2(y) + 2 \ddot{a}(y)
+ m^2 \exp[2 a(y)] \r)Z = 0.
\label{app:schr1}
\e

\end{document}